# Electromagnetic Field Tapering in the High-Roughness Substrates Coated by a Thin Film of Manganese: A Lithography-Free Approach to Ultra-Broadband, Wide-Angle, UV to FIR Perfect Absorption

Majid Aalizadeh,[1,2,*] Mohammad Reza Tavakol,[3] Amin Khavasi,[3] Mehmet Yilmaz,[4,5] Ekmel Ozbay[1,2,5,6]

[1] Department of Electrical and Electronics Engineering, Bilkent University, Ankara 06800, Turkey
[2] Nanotechnology Research Center (NANOTAM), Bilkent University, Ankara 06800, Turkey
[3] Electrical Engineering Department, Sharif University of Technology, Tehran, 11155-4363, Iran
[4] Institute of Materials Science and Nanotechnology, Bilkent University, Ankara, 06800, Turkey
[5] UNAM - National Nanotechnology Research Center, Bilkent University, Ankara, 06800, Turkey
[6] Department of Physics, Bilkent University, Ankara 06800, Turkey
*majid.aalizadeh@bilkent.edu.tr

## Abstract

Metallic layers are known to be used for the suppression of wave transmission when their thickness is sufficiently higher than the skin depth of metal. If in addition to blocking the transmission, metallic layers have the feature of blocking the reflection, too, they would make perfect absorbers. In this work, we propose an experimental approach of using a single thin layer of Manganese (Mn) as both the transmission suppresser and the reflection suppresser. This approach leads to obtaining lithography-free ultra-broadband perfect absorption in an ultra-wide spectrum ranging from Ultraviolet (UV) to Far Infrared (FIR). The measured average absorption is approximately 99%. Such a promising result can be achieved by only coating a single Mn layer on high-roughness substrates that include random nano-pyramids on it. In other words, we do not need a stack of different materials and combinations of geometrical features. The high roughness is realized on a commercial Silicon wafer substrate by inductively coupled plasma (ICP) etching. The key to this ultra-wideband absorption is electromagnetic field tapering which exists due to the graded-index feature of the structure (known as moth-eye effect), along with the ideal optical properties of Mn which makes it an excellent metal for broadband absorption applications. A full experimental characterization of the fabricated samples is presented along with the physical analysis of the phenomena. The findings of this paper can be used for the realization of lithography-free, cost-effective and high-throughput mass production of broadband absorbers.

## Introduction

Broadband absorbers are increasingly attracting attention, because of their wide range of applications including photovoltaics [1-4], thermal emission [5], photodetection [6-10], thermal imaging [11-13], shielding [14,15], etc. Metamaterials have made it possible to create subwavelength broadband absorbers that meet the requirement of downscaling trend of optical devices. In such cases, generally the structure consists of a thick bottom metallic layer to ensure zero transmission, and the rest of the structure comprising couple of layers, with different materials and sometimes patterns, play the role of suppression of reflection. Blocking reflection is obtained by designing the structure to have its impedance matched to that of the free-space [16].

The typical ideas for achieving broadband absorption are decreasing the quality factor of the resonance mode contributing to the absorption, or combining different adjacent resonances to build a broadband multi-resonance response which can be qualitatively described as the superposition of the consisting resonances.

As an example for the first approach, i.e. decreasing the quality factor, we can mention decreasing the quality factor of the Fabry-Perot cavity resonators. This happens, for example, by using highly lossy metals such as Mn [17] or Chromium (Cr) [18], in the metal-insulator-metal configuration which provides a wideband impedance matching with free-space, leading to broadband absorption. Decreasing the quality factor is also obtainable by using metal-insulator-metal-insulator configuration where the top insulator is added as an intermediator layer to make the impedance transition from air to the cavity more gradual [19,20]. This eventually leads to broadening the impedance matching band.

About the second approach which is taking the advantage of the superposition of adjacent resonances, we may refer to one of the pioneer works carried out by Aydin. et al [21]. In that work, they adopted the collective absorption effects of adjacent Localized Surface Plasmon (LSP) resonances caused by varying widths of metallic nanorods and obtained an average absorption of 71% in the range of 400-700 nm. Another example can be superposition of LSP resonances induced by randomly distributed nanoparticles on the top layer of an MIM configuration which leads to high absorption in the visible regime [22].

One huge barrier toward mass, large-area and high-throughput production of broadband absorbers is the costly and time-consuming stage of lithography. The typical used type of lithography is Electron Beam lithography (EBL). For the purpose of intuition, patterning a 1 $cm^2$ area of sample may take up to a several days. Therefore, this is a crucial limitation. As a result, coming up with lithography-free structures is of high importance for large-scale production [23].

Some of the implemented ideas for lithography-free broadband absorbers in previous works include Pt-coated randomly distributed chemically synthesized dielectric nanowires [24], high-temperature annealing of thin metallic layers to deform them into random nanoparticles or nanoholes [22,25,26], typical MIM or MIMI cavity-based absorbers [19], and random nanopillars [27].

In this work, we present a facile and straightforward approach for the fabrication of ultra-broadband lithography-free absorbers by a single film of Mn coated over a high-roughness substrate including randomly sized needle-like and sharp nano-pyramids. The measured absorption using an integrating sphere confirms an average of 99% absorption in the ultra-broad range of 250-2500 nm, i.e. from UV to MIR. Simulations predict that the absorption above 90% regime should at least go above 10 microns, which means UV to FIR. It is explained that such a broadband absorption is due to the cooperative contribution of two key factors: 1. electromagnetic field tapering owing to the tapered nature of the structure (graded-index or moth-eye effect), and the very appropriate optical properties of Mn for broadband absorbers [28,29]. These physical characteristics contribute to the fact that the coated Mn film surprisingly, not only blocks the transmission, but also suppresses the reflection. Such a result, to the best of our knowledge, is the best absorber realized with lithography-free structures. This concept can be generalized to high-roughness substrates with different materials, as well, because as mentioned, the field does not penetrate below the Mn layer and therefore, the substrate only is required to have tapered features on it, independent of the type of its material. Such a desirable result with a relatively simple and inexpensive fabrication process, can be an excellent candidate for mass production purposes with a variety of applications ranging from photovoltaics to thermal emission and shielding.

# Result and Discussion

Planar metal films with a thickness sufficiently larger than their skin depth are used as the bottommost part of the broadband absorbers to guarantee almost zero transmission. However, the same metal film if used solely, leads to a high reflection due to its large impedance mismatch with free-space. Therefore, absorbers typically include other layers above the metallic substrate that suppress the reflection by providing impedance matching with the free space. In this work, we use only one metallic thin film of the thickness of 600 nm coated on a substrate with tapered features to obtain transmission and reflection suppression at the same time in an ultra-broad range, and thus reach an ultra-wideband absorption. This design does not include any other layers or materials on top of this metallic film.

Figure 1(a) shows the plots of transmission (T), reflection (R), and absorption (A) for a 600 nm planar thin film of Mn. The absorption can be calculated as A=1-R-T. It can be observed that as expected, a metallic layer with such a thickness leads to almost zero transmission, however, the reflection is large, and thus the total absorption is low. The idea in this work is to coat the same type of metallic single thin film but to make it in a tapered configuration that enables the moth-eye or graded-index effect. Figure 1(b) shows the simulated R, T, and A plots for the case of making the thin film of Mn in a periodic pyramidal configuration with the thickness of 600 nm, where the period is 1 micron and the height of pyramids is 4 microns. In other words, we have suspended periodic pyramids made of 600 nm thick Mn layer. The schematic of such a case is shown in the inset of Fig. 2(b). We note that by making this thin film tapered, in addition to the suppression of T, R gets suppressed in an ultra-broad range of wavelenghs. This leads to having perfect absorption in the broad band of UV to FIR. The physics behind this broadband absorption will be elevated in detail in the next sections. It is also noteworthy that the permittivity of Mn (and Si) used throughout the paper and in the simulations is measured by fitting measured spectroscopy data obtained by J.A. Woollam Co. Inc. V-VASE and IR-VASE Ellipsometers.

In order to experimentally implement the abovementioned idea, we should have a substrate with tapered features on it to coat the Mn layer. For instance, one can pattern a substrate with the shape of periodic nano-pyramids using EBL or FIB, but these methods are expensive and very time-consuming, and they do not provide us with large-area structures. To avoid such a limitation, we use a lithography-free method which leads to randomly-sized nano-pyramids on the surface of a commercial 4-inch p-type Silicon wafer. It is carried out by using inductively coupled plasma (ICP) etching. The use of etching to obtain nano-pyramids on the surface of a Si wafer is carried out in the previous works with other recipes, and leads to the so-called Black Silicon which has applications as broadband absorber or solar cell. However, as will be experimentally presented in the latter sections, its absorption is limited up to around 1000 nm because of the limited absorption properties of the doped Silicon. However, as will be shown in the next sections, any substrate made of any other material that has tapered features on it gives the same result and the ultra-broadband absorption of the structure is not due to the properties of Si and is because of the tapered geometry and optical properties of the Mn layer.

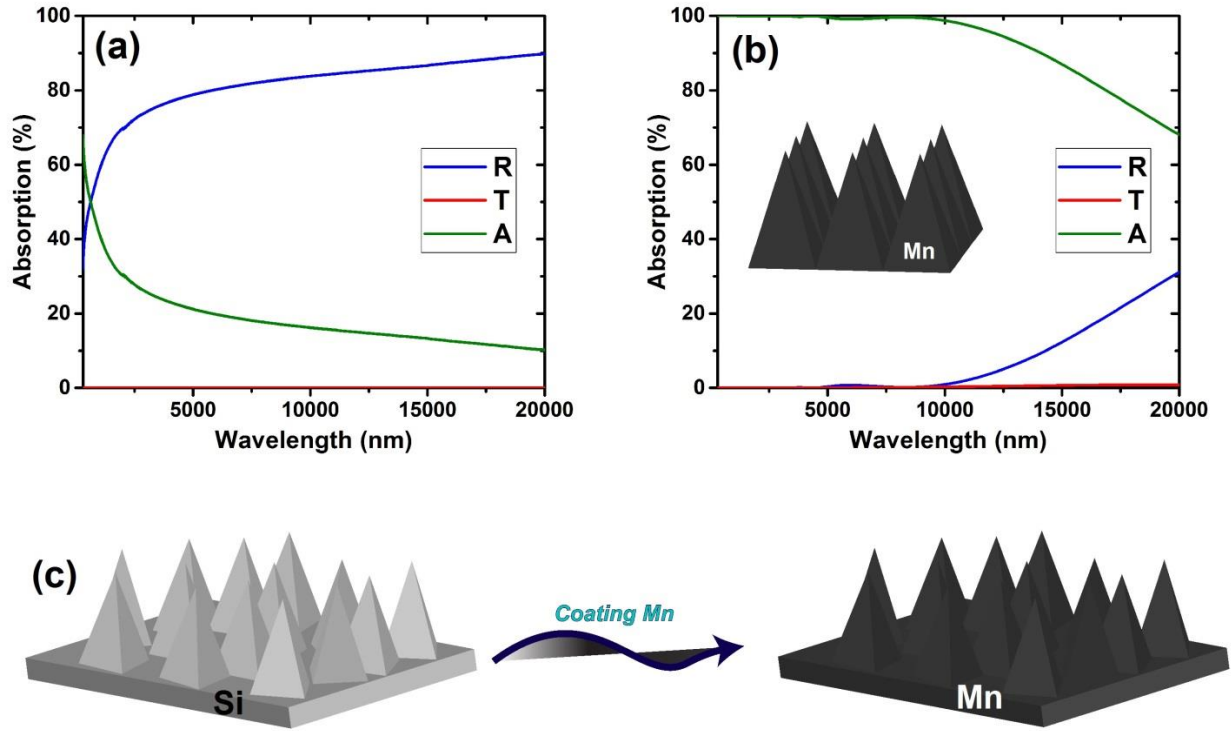

Figure 1. a) Transmission, reflection, and absorption spectrum of a 600 nm thick planar Mn layer, b) Transmission, reflection, and absorption spectrum of a 600 nm thick suspended pyramidal array of Mn, and c) schematic of the random Si nano-pyramids before and after coating of Mn.

Figure 1(c) shows the schematic of the randomly sized nanopyramid array on the surface of a Si wafer before and after coating Mn. In order to achieve such features on the surface of the Si substrate we have adopted a recipe of ICP with cyclic steps, using STS Multiplex ICP Etch system. Each cyclic step consists of two half steps. The first half step is reactive ion etching (RIE) of Si and the second half step is deposition of $C_4F_8$ on RIE etched Si surface. In the first half step (RIE) a mixture of $SF_6$ and $O_2$ gasses are released into the process chamber for duration of 10 seconds. In the second half step (deposition), $C_4F_8$ is deposited on the Si surface for 7 seconds. Gas flow rates for $SF_6$, $O_2$, and $C_4F_8$ gases are set as 80, 5, and 70 sccm, respectively. The chamber pressures during the half steps of etching and deposition are set as 35 and 20 mTorr, respectively. During the etching and deposition half steps, the coil powers are set as 500 and 400 Watts, respectively, and the platen powers are set as 13 and 0 Watts, respectively. The chamber temperature is set to 20°C for both of the half steps, and hence the entire process. Table 1 shows the ICP recipe parameters for both deposition and etching half steps.

| Parameter | Etching | Deposition |
|---|---|---|
| duration time (s) | 10 | 7 |
| $SF_6$ flow rate (sccm) | 80 | |
| $O_2$ flow rate (sccm) | 5 | |
| $C_4F_8$ flow rate (sccm) | | 70 |
| chamber pressure (mTorr) | 35 | 20 |
| coil power (Watts) | 500 | 400 |
| platen power (Watts) | 13 | 0 |
| chamber temperature (°C) | 20 | 20 |

Table 1. ICP recipe parameters for both deposition and etching half steps of Silicon wafer.

Figure 2(a-d) demonstrate the Scanning Electron Microscopy (SEM) images of the surface of a commercial p-type silicon wafer substrate when 100, 200, 300, and 400 cycles of the abovementioned ICP recipe is applied on it, respectively. It can be observed that with 100 cycles of etching, tip of the random nano-pyramids start to appear, however, there are still significant flat spaces on the surface. For the cases of 200, 300, and 400 etching cycles, the nano-pyramids have obviously appeared and as expected, the height of pyramids increases as the number of cycles increase. In other words, the etching depth increases with the number of cycles. It can be observed in Figs. 2(b-d) that the shaped pyramids are sharp and needle-like and they randomly vary in their size and distribution. The flat areas without tapered features on the surface of the 100-cycle etched wafer make it inappropriate for our purpose. Therefore, we coat Mn on the surface of the other three samples with 200, 300, and 400 etching cycles. Using the abovementioned recipe, we obtained such features on a full 4-inch Si wafer, therefore, in principle, this process is wafer-scale and can be applied to any size of wafers. However, the process parameters may need to be modified if larger scale wafers (i.e. 6-inch diameter, or 8-inch diameter Si wafers), or smaller scale wafers (i.e. 2-inch diameter Si wafers) are used instead of 4-inch diameter Si wafers.

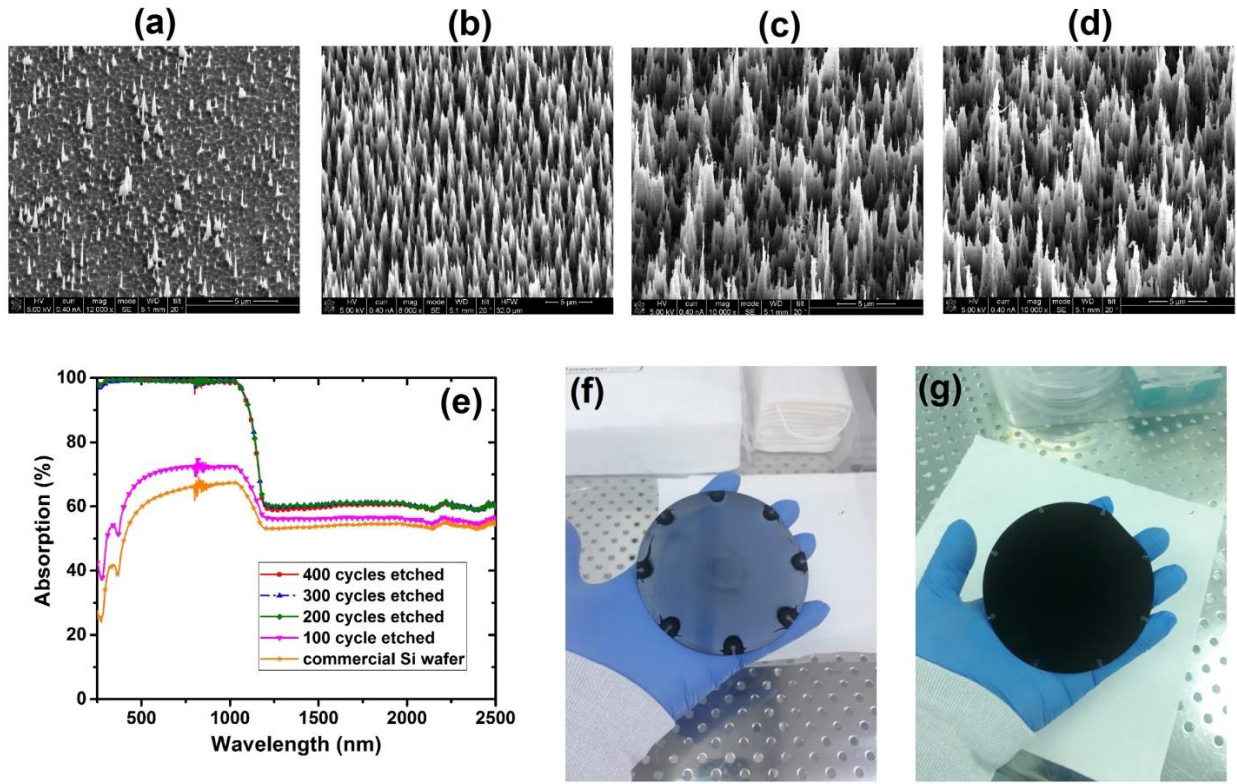

Figure 2. SEM image from the surface of the Si wafer etched for a) 100, b) 200, c) 300, and d) 400 cycles. e) Absorption spectra of the samples mentioned in (a), along with commercial Si wafer. Photograph of the Si wafers etched for f) 100, and g) 400 cycles.

The measured absorption spectra of the samples, the SEM images of which are shown in Figs. 2(a-d) are demonstrated in Fig. 2 (e), along with the absorption spectrum of the commercial Si wafer. It can be observed that as expected, the 100-cycle sample does not give high absorption and has an absorption spectrum close to that of Si wafer before etching. On the other hand, all other three samples with 200,

300, and 400 etching cycles show very high absorption in the range up to around 1000 nm wavelength, and their absorption profiles are very similar to each other. The absorption of these samples drops significantly after 1000 nm which is due to the reduction in the loss of the Si wafer. The images of the 100-cycle and 400-cycle etched Si full wafers are shown in Figs. 2(f) and 2(g), respectively. The black color of the 400-cycle sample is attributed to the high absorption of the nano-pyramids of doped Si in the visible region. However, as seen in Fig. 2(e), the absorption in this case is limited up to around 1000 nm wavelength, and for larger wavelengths, the structure loses its high absorption behavior. On the other hand, the 100-cycle sample has a non-black color close to that of commercial wafer without etching, because of its weak absorption.

Here, we emphasize that in order to obtain the absorption (A) spectrum, we should measure the transmission (T) and reflection (R). T can be measured by using normal devices such as ellipsometer or FTIR. However, for measuring R for this structure, we require an integrating sphere set up to be integrated with the device we are using. This is due to the fact that because of the nonplanar geometry of the structure, a high portion of the reflected wave will be scattered, i.e. we have to measure both diffusive and specular reflections from the sample. Any device without an integrating sphere setup only measures the specular reflection. However, the integrating sphere has a reflective mirror inside it and gathers all portions of the reflected beam which is scattered to any direction and sends all to the detector. The integrating sphere used in our device, is attached to the Carry 5000 UV-VIS-IR spectrophotometer and measures in the range of 250-2500 nm. The sphere is covered with PTFE and it has a 110 mm diameter. The transmission measurements are carried out using V-VASE and IR-VASE Ellipsometers.

In order to extensively broaden the perfect absorption band, we coat a single layer of Mn with three different thicknesses of 400, 600, and 800 nm, using VAKSIS electron-beam evaporator system. The coating rate of Mn is set to 10 A/s. The relatively high rate of coating is due to the fact that in this work it is not required to have flat and conformal films, therefore, we can use high coating rates which is very desirable for high-throughput production purposes. On the other hand, in broadband absorption applications, having extra roughness and non-conformal layers generally leads to a broader absorption band since these features may add extra resonances to the overall performance of the structure.

The SEM image of the 300-cycle sample coated with 600 nm Mn is shown in Fig. 3(a). Moreover, Fig. 3(b) demonstrates the cross-section SEM image of the same sample which is milled by Focused Ion Beam (FIB). Fig. 3(c) also shows the Atomic Force Micrcoscopy (AFM) image of the surface topography of the sample. It can be observed that the randomly sized Si nano-pyramids are coated with Mn and the Mn layer's thickness is not uniform. Finally, Fig. 3(d) shows the image of the mentioned sample.

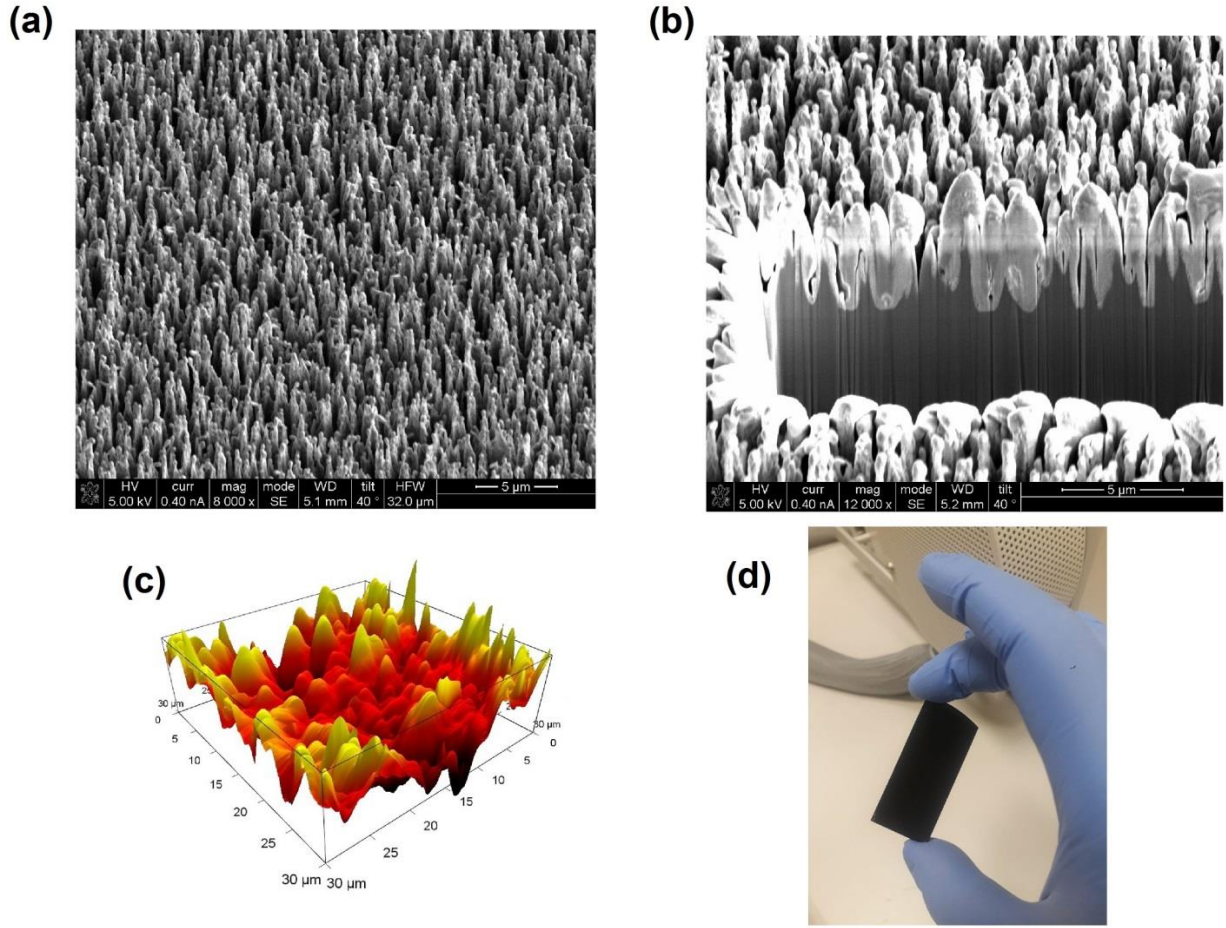

Figure 3. a) SEM image of the surface of the 300-cycle etched sample coated with 600 nm Mn, b) SEM image of the cross-section of the sample mentioned in (a) that is milled by FIB, c) AFM image of the surface topography of the sample mentioned in (a), and d) photograph of the sample mentioned in (a).

Figures 4(a), (b), and (c) show the absorption spectrum for the 200, 300, and 400 cycles etched samples, respectively. Each figure contains the measured absorption for the three thicknesses of coated Mn, 400, 600, and 800 nm. The variation of etching cycles and coated Mn thickness is to check the possible changes of the absorption spectrum's behavior with regards to the fabrication parameters. It can be observed that throughout the whole measured wavelength range of 250-2500 nm, the absorption is almost perfect for all fabricated devices, and their performances are almost the same. In other words, the measured absorption confirms perfect absorption ranging from UV to MIR with negligible dependence to variations in the fabrication process, and the structure shows an excellent fabrication tolerance. The average absorption in the measured range is approximately 99%. Such wideband absorption with a lithography-free structure has not been experimentally reported before, to the best of our knowledge. It should be also noted that the integrating sphere setup is limited up to the wavelength of 2500 nm. As it will be shown in the followings, simulations confirm that the absorption band of the structure goes up to around 10 microns, i.e. from UV to FIR. It is also evident from the trend of measured absorption behavior of the structure in Fig. 3 as the wavelength increases, that the absorption does not show a negative slope toward the larger wavelengths, and if the same trend continues for the larger wavelengths that have not been measured, the absorption will remain above 90% for up to a couple of microns larger than 2500 nm wavelength. It is noteworthy here that as it will be confirmed by the simulations, the high-roughness Si

wafer does not contribute to blocking the transmission, and is it almost completely blocked by the Mn layer which in all cases is couple of hundreds of nanometers. The color of 600 nm Mn-coated 300-cycle sample shown in Fig. 3(d) as well as all other samples is black due to the perfect absorption of the visible light. The objective of fabricating 9 samples with 3 different etching cycles and 3 different Mn thicknesses is to show that this method has a very high fabrication tolerance and the combination of tapered substrate with the optical properties of Mn is the key to this broadband absorption and it is not restricted to any specific geometrical optimization. This is a very important feature for mass production of such absorbers.

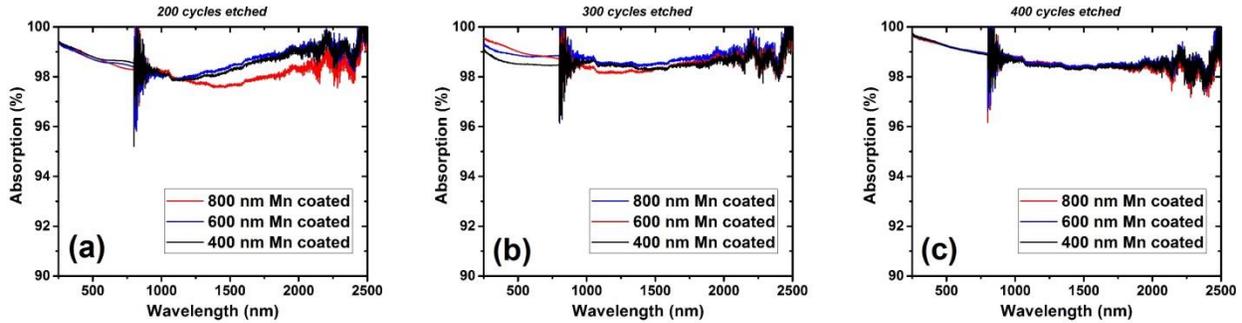

Figure 4. Absorption spectra of the fabricated samples with the three different Mn coating thicknesses of 400, 600, and 800 nm, for the cases of a) 200, b) 300, and c) 400 cycles of etching.

Based on the topographic image of the sample (Fig. 3(c)) we can approximately model the 300-cycle 600 nm Mn-coated structure as a periodic array of Mn-coated pyramids, the xz cross-section of the unit cell of which is shown in Fig. 5(a). In this approximation, the whole structure is assumed to be made of similar periodic pyramids. As can be seen in Fig. 5(a), the height of the inner pyramid made of Si, the period, and the thickness of the coated metal layer are denoted as h, p, and t, respectively. Based on the SEM images taken from the cross-section of the sample, we choose h=4 microns, p=1 micron, and t=600 nm for the simulation. In this unit cell, the Si pyramid has not extended downwards below the region where Mn is coated. This is done to make sure that the transmission blockage behavior of the structure is not due to the high thickness of the Si wafer below the patterns, and is only achieved by the Mn layer.

The absorption of the approximated structure is shown in Fig. 5(b). The simulations are carried out with two different methods of FDTD and FEM using the Lumerical FDTD and COMSOL Multiphysics softwares, respectively. It can be observed that such a structure shows absorption of above 90% up to the wavelength of 13500 nm, i.e. spanning from UV to FIR. To discuss the results more clearly, the simulated reflection and transmission of this sample are also shown in the same plot. It can be observed that in the mentioned ultra-broad range, both the reflection and transmission are suppressed. The transmission is suppressed due to the fact that the thickness of the Mn layer is 600 nm which is much larger than its skin depth, therefore, this simulation proves that in the experimental case, the transmission is not blocked as an effect of high thickness of the Si wafer below the nano-pyramids and is a direct result of the Mn layer's thickness solely. This also enables us to fabricate the same structure on top of any high-roughness substrate other than Si, and shows that the absorption behavior of the structure is independent from the choice of substrate. We further demonstrate this point by studying the field profile within the structure.

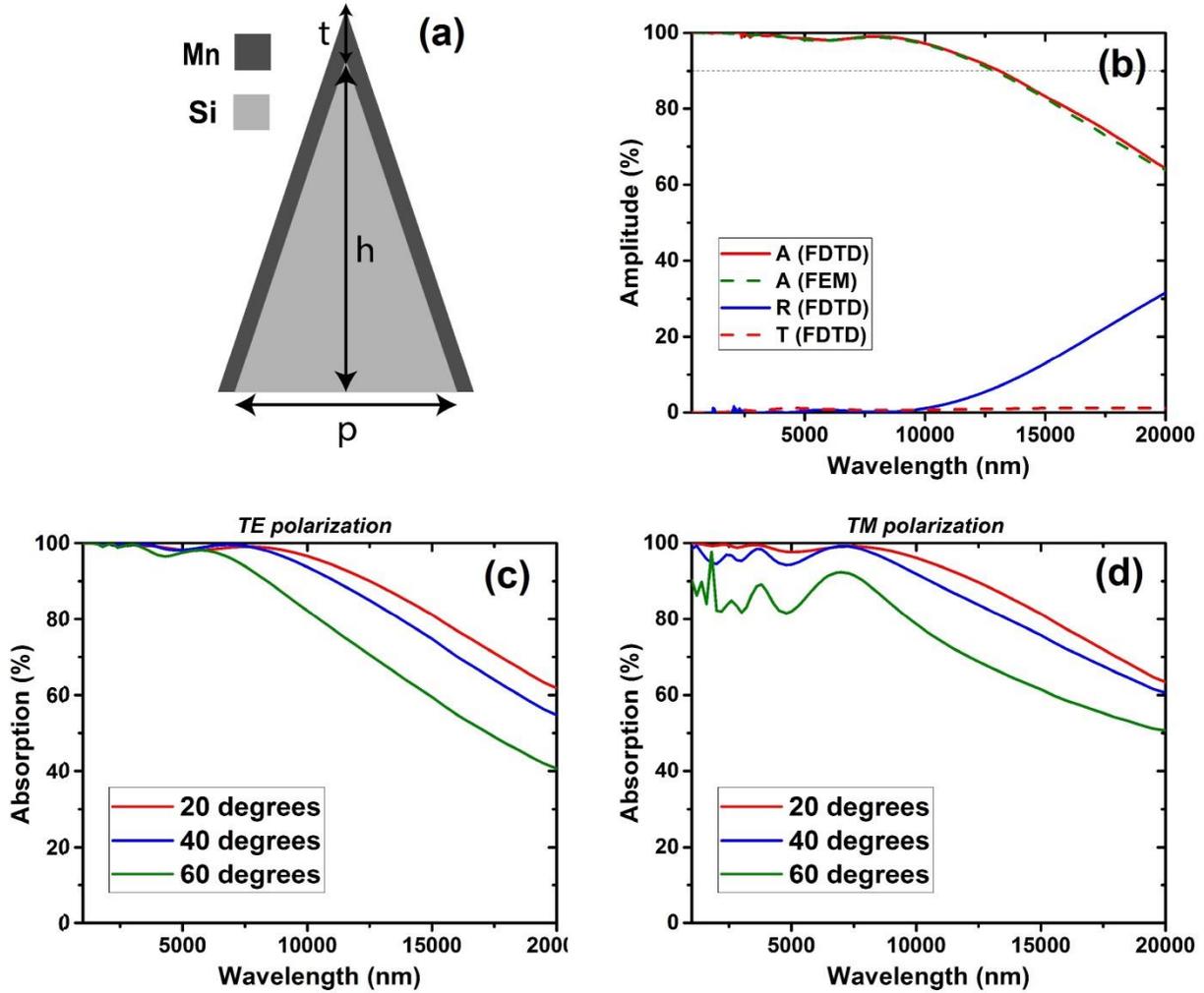

Figure 5. a) xz cross-section schematic of the unit cell of the approximated model of the structure, b) Calculated reflection, transmission, and absorption spectrum of the structure when h=4, p=1, and t=0.6 microns. Calculated absorption spectra for the the incidence angles of 20, 40, and 60 degrees, for the case of c) TE, and d) TM polarization.

We note that the reflection is suppressed because of the effect of Field Tapering or the graded-index (moth-eye) effect. So, unlike conventional structures in which the metallic layers of such a thickness are put at the bottommost part to only block the transmission, and the minimization of reflection happens due to the contribution of the other parts of the structure above that bottom layer, in our case, a single patterned metallic layer happens to play the role of blocking both the transmission and reflection at the same time. It is due to the tapered geometric shape of the substrate which enables the effect of field tapering. This effect, along with the very appropriate optical properties of Mn and the light trapping nature of this structure, which will be discussed in detail later, contributes to such a broadband absorption. Since it is not possible to use integrating sphere setup for the case of oblique incidence, the same approximated structure is also simulated for the cases of TM and TE polarizations of oblique incidence. The absorption spectra for incidence angles of 20, 40, and 60 degrees and for TE and TM polarizations are presented in Figs. 5(c) and 5(d), respectively. It can be observed that the structure retains its high absorption for large incidence angles and for both polarizations although it is more robust for the case of TE polarization.

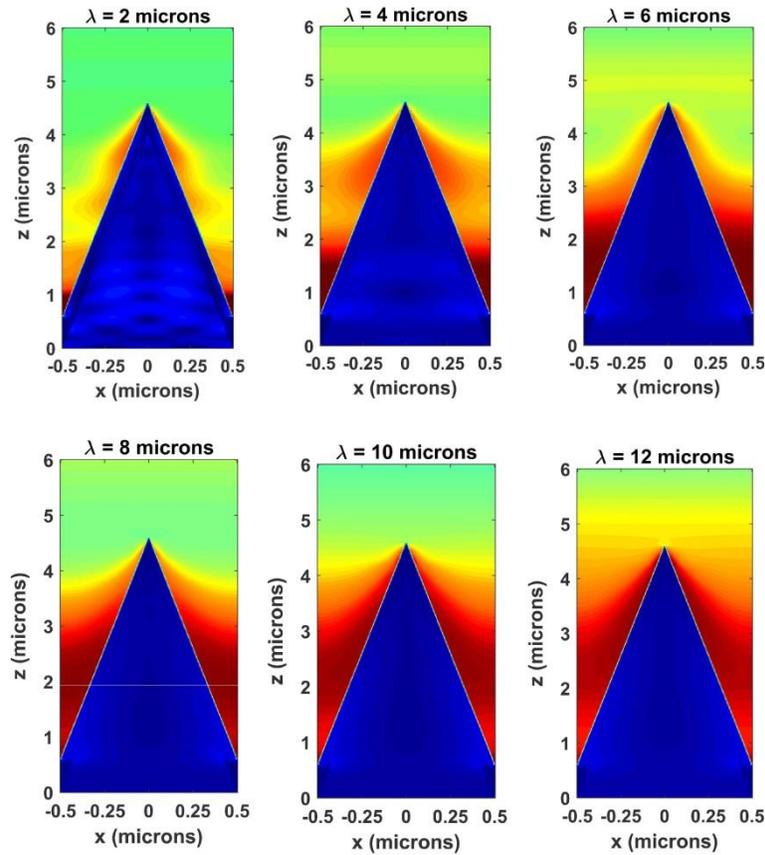

Figure 6. Electric field magnitude distribution at the xz cross-section of the unit cell for different wavelengths of 2, 4, 6, 8, 10, 12 microns.

Now let us discuss the physical mechanism behind this broadband absorption in more detail. As the first step toward the physical analysis, the electric field magnitude pattern is shown at the xz-plane cross-section of the unit cell in Fig. 6. The simulations are carried out for the different wavelengths of 2, 4, 6, 8, 10 and 12 microns. The field patterns clearly demonstrate the phenomena of the field tapering due to the graded-index nature of the structure which is also referred to as the moth-eye effect. Field tapering happens when there is a gradual change in the effective index of the structure from the top to the bottom which leads to a slow and smooth impedance matching between the air and structure. Here obviously the structure-to-air ratio at any horizontal cross-section of the structure increases gradually which is the cause of graded-index effect in this pattern. This shows that the incident field slowly gets coupled to the structure in the region between the adjacent unit cells, but it cannot penetrate inside the coated pyramids due to the high thickness of the Mn layer. After coupling slowly to the structure, the field gets absorbed by the lossy Mn layer.

Here we demonstrate why Mn is an appropriate metal for this structure. Figure 7(a) shows the absorption spectrum for different coated metals in the configuration of Fig. 5(a), which are typically used for broadband absorber structures. These metals include W, Cr, Ti, Fe, and Mn. The dimensions are as chosen before. It can be seen that for the case of other metals, the A>90% band at the best goes up to the wavelength of approximately 4.5 microns for the case of Ti, while when using Mn the bandwidth enhances very significantly and the upper wavelength limit of A>90% band goes up to around 13 microns. This significant superiority of Mn can be explored in its optical properties. For this purpose, we have plotted the real part of relative permittivity of all the metals in Fig. 7(b). The optical data for Mn is obtained by ellipsometric measurements and for all other metals it is taken from Palik. It can be observed that in such a broad wavelength range, Mn has a much smaller absolute value of real part of permittivity compared to the other metals and it retains its small value throughout the whole spectrum. Having a small real part of permittivity leads to better field penetration inside the metal and through the intrinsic loss of metal, i.e. the imaginary part of permittivity, the penetrated field gets absorbed. Moreover, the absorption spectra of 600 nm planar films of the abovementioned metals are shown in Fig. 7(c). It can be observed that even a planar film of Mn has a better absorptive capability compared to other metals.

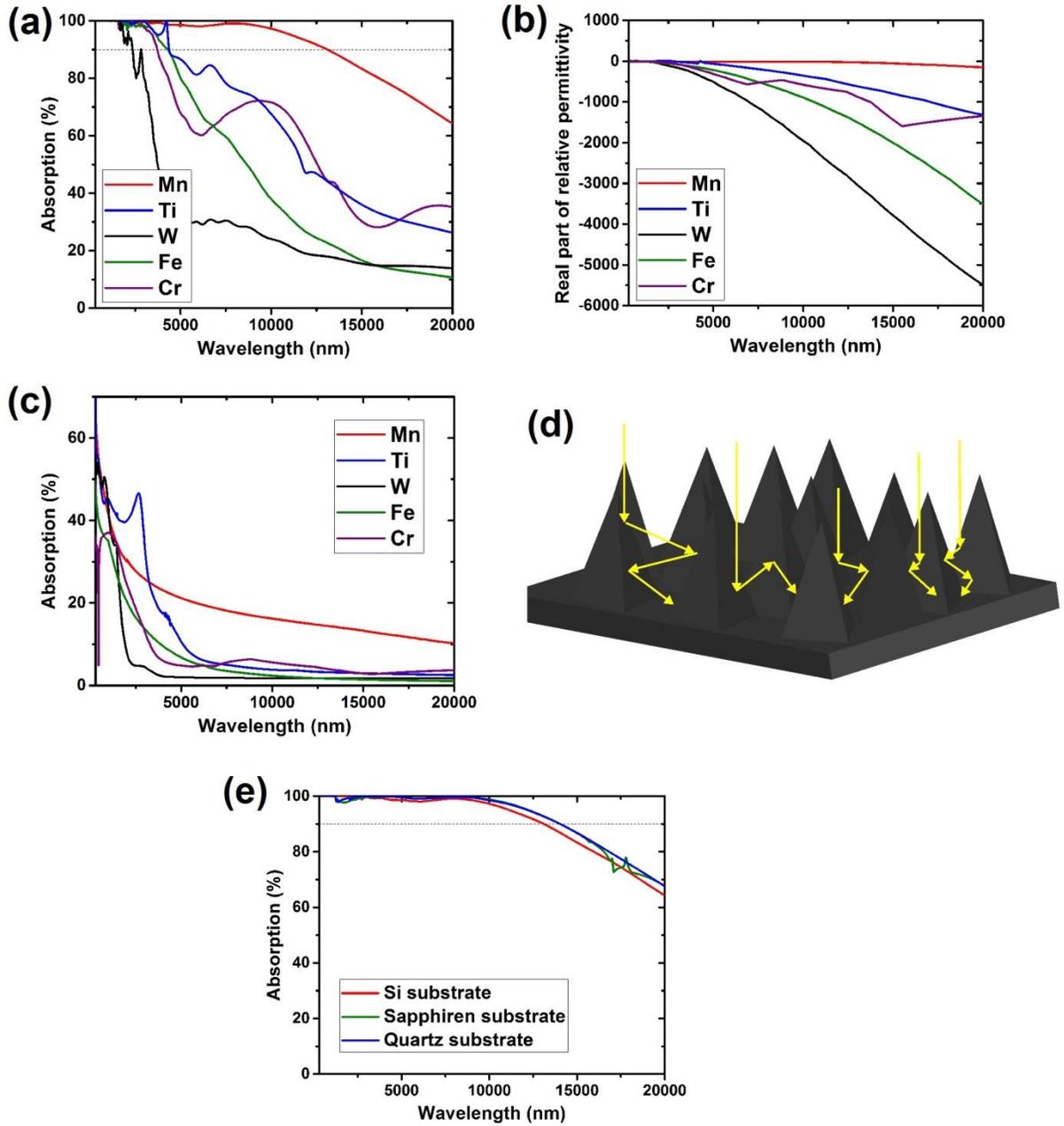

Figure 7. a) Absorption spectrum of Si pyramid array coated with, b) real part of relative permittivity, and c) absorption spectrum of 600 nm thick planar film of: Mn, Ti, W, Fe, and Cr. d) Schematic of the light trapping nature of the structure, and, e) absorption spectrum of the pyramid array of different substrates when coated with Mn.

The other physical mechanism (especially in shorter wavelength where ray optics approximation is applicable) contributing to the excellent performance of the proposed device is the light-trapping nature of the structure. It is common to trap the light inside fabry-perot or gap-plasmon cavity modes, but in this structure, the nonplanar configuration of the constituting elements makes it a nice scaffold for light trapping. As shown by the schematic presentation in Fig. 7(d), the light can get reflected back and forth between the neighboring pyramids and in each reflection it can be partially absorbed.

Based on the abovementioned arguments it can be summarized that the main physical mechanisms behind this broadband strong absorption are: 1. The phenomena of field tapering which exists due to the graded-index profile of the structure, and, 2. The significantly small real part of permittivity of Mn, which allows better field penetration inside it and consequently leads to higher absorption compared to conventional absorbing metals. 3- The special geometry of the structure which makes it an appropriate platform for light trapping.

As a final remark, in order to show that this approach is valid to any high roughness substrate and it is not a result of Si properties, we have shown in Fig. 7(e) the absorption spectrum for the case of using the unit cell shown in Fig. 5(a), but with different substrates. The other possible typical substrates are quartz ($SiO_2$), and Sapphire ($Al_2O_3$). It is obvious that the result is almost robust to the choice of substrate and the ultra-broadband absorption profile is almost the same for all the substrate materials.

## Conclusion

In summary, a facile and feasible lithography-free approach for obtaining ultra-broadband perfect absorption through a single layer coating of Manganese over a high-roughness substrate including random nano-pyramids, is experimentally presented. The experimental measurements using integrating sphere confirm approximately 99% average absorption in the range of 250 to 2500 nm, i.e. from UV to MIR. The simulations predict that the absorption band extends up to above 10 microns, in other words, from UV to FIR. The physical mechanism behind this broadband absorption is discussed and it is shown that this is due to the cooperative contribution of superior optical properties of Mn along with the phenomena of electromagnetic field tapering and light-trapping between the pyramids. The tapered structures on the surface of the Si substrate are obtained by ICP etching. It is important to emphasize that, as demonstrated through the simulations, this approach can be used for any high-roughness substrate other than Silicon with pyramidal-shaped features on it. By fabricating several samples with different etching cycles and Mn layer thicknesses, and by achieving almost the same result from all of them, it is also experimentally shown that the structure has a high fabrication tolerance which makes it a very appropriate candidate for mass production. The findings and experiments of this paper are very beneficial for large-area and mass production of broadband absorbers covering the full range of UV to FIR.